\documentclass[preprint,pteplogo]{ptephy_v2}
\usepackage{hyperref}
\usepackage[T1]{fontenc} 
\usepackage{slashed}
\usepackage{here}

\newcommand{\tr}{\mathop{\rm Tr}}
\def\Tr{\text{Tr}} 
\def\nn{\nonumber}

\begin{document}

\title{\boldmath Violation of non-Abelian Bianchi identity and QCD topology}
\author{Tsuneo Suzuki}
\affil{Kanazawa University, Kanazawa 920-1192, Japan \email{suzuki04@staff.kanazawa-u.ac.jp}} 

\begin{abstract}%
If a gauge field in QCD has a singularity of the Dirac type, the non-Abelian Bianchi identity is violated. The violation of the non-Abelian Bianchi identity (VNABI) $J_{\mu}(x)$ is equal to 8 degenerate Abelian magnetic monopole currents $k^a_{\mu}(x)\ \ (a=1\sim 8)$.  Abelian monopoles could exist without any artificial Abelian projection.  When $J_{\mu}$ condenses in the vacuum 
as suggested in latttice QCD simulations, 
color confinement of QCD is realized by the Abelian dual Meissner effect (ADME).   
It is important to study the effect of VNABI besides the role of color confinement mechansim.
It is found that
VNABI affects topological features of QCD. Firstly,  the topological charge density $\rho_t(x)=\Tr G_{\mu\nu}(x)G^*_{\mu\nu}(x)$ is not expressed by a total derivative of the Chern-Simons density $K_{\mu}(x)$, but has an additional term $L(x)= 2\Tr(J_{\mu}(x)A_\mu(x))$, that is, $\partial_\mu K_\mu(x)=(g^2/16\pi^2)(\rho_t(x)-L(x))$. Secondly, the axial $U(1)$ anomaly is similarly modified  as $\partial_{\mu}j_{\mu}^5(x)=2m\bar{\psi}(x)\gamma_5\psi(x)+(g^2/8\pi^2)(\rho_t(x)-L(x))$ while the Atiyah-Singer index theorem  is formally unchanged. 
However, the integrated term $\Lambda=(g^2/16\pi^2)\int d^4xL(x)$ is seemingly not integer nor gauge-invariant. Hence if $\Lambda$ remains non-zero, VNABI is not allowed in QCD. 
  Abelian monopoles as VNABI are regarded as a generalization of the Dirac monopole to a $U(1)$ subgroup in non-Abelian gauge theories as discussed  by Wu-Yang. When we make use of  similar arguments, the additional term $\Lambda$  is proved to vanish theoretically.  We also evaluate
the term $\Lambda$ in the framework of Monte-Carlo simulations on $24^4$ lattices at the lattice spacing between $0.05\sim 0.17$fm adopting an tadpole-improved $SU(2)$ gluonic action. To reduce ultraviolet fluctuations, we introduce a partial gauge fixing like the Maximal Center gauge (MCG). 
 The term $\Lambda$ is largely fluctuating around zero and $|\Lambda|$ is decreasing as $\beta$ becomes larger. Then the gradient flow method is applied. The term $\Lambda$ 
tends to vanish rapidly after small gradient flow time 
$t_{flow}$.
This suggests that the term $\Lambda$  becomes zero in the continuum limit. 
Hence VNABI could exist in QCD.
The biggest effect of VNABI on QCD topology seems to be that 
self-dual instantons can not be a classical solution of QCD 
at space-time points where VNABI occurs. One has to find an alternative mechanism explaining integer topological charge, etc.
A new interesting relation is derived with respect to the topological charge. When $\Lambda=0$, an Abelian counter term $Q_a\equiv (g^2/16\pi^2)\int d^4x \Tr(f_{\mu\nu}f_{\mu\nu}^*)$ written by Abelian field strengths is proved theoretically to satisfy $Q_a$=3$Q_t$ where $Q_t=(g^2/16\pi^2)\int d^4x \rho_t(x)$. 
The fact sugggests that the Abelian magnetic and electric fields in the presence of $J_{\mu}$ condensation could have a role in explaining the topological charge 
$Q_t$ in place of instantons. 
\end{abstract}


\maketitle
\section{Introducation}
    In non-perturbative QCD, color confinement and chiral symmetry breaking are two important unsolved problems. Topological objects such as monopoles and instantons are believed to play an important role in these phenomena.     
The mechanisms of the phenomena and the relation between these topological quantities have been unclear. 

  Mandelstam~\cite{Mandelstam:1974pi} and 't Hooft~\cite{tHooft:1975pu} proposed a dual picture of the Meissner effect as the color confinement mechanism~\cite{Nambu:1974} assuming that any topological magnetic excitation existing in QCD condenses in the vacuum.    
  An exact approach to Abelian magnetic monopoles in QCD was made by 't Hooft~\cite{tHooft:1981ht} who introduced a partial gauge fixing $V(x)$ of $SU(3)$ keeping  the maximal torus group $U(1)\times U(1)$ unbroken. Although there are many numerical results suggesting color confinement due to the Abelian dual Meissner effect (ADME)~\cite{Suzuki:1990,Kronfeld:1987ri,Kronfeld:1987vd,Shiba:1994,Stack:1994,Koma:2003B,Sekido:2007,Sakumichi:2014}, the 't Hooft scheme is dependent on the choice of the gauge fixing matrix  $V(x)$~\cite{Maxim:2004}
and not satisfactory at least on lattice.

With respect to the gauge dependence problem of the Abelian projection,  Bonati et al.~\cite{Bonati:2010tz} proposed an interesting possibility that violation of the non-Abelian Bianchi identity (VNABI) exists behind any type of the Abelian monopole from the Abelian projection~\cite{tHooft:1981ht}. Although they expected the gauge independence of the Abelian projection scheme~\cite{tHooft:1981ht}, it is not correct as clarified below.

     Motivated by the work~\cite{Bonati:2010tz}, the present author~\cite{Suzuki:2014wya,Suzuki:2017lco} found a new interesting general fact. Define a covariant derivative operator  as $D_{\mu}=\partial_{\mu}-igA_{\mu}(x)$ and a non-Abelian field strength as $G_{\mu\nu}$. The Jacobi identities are expressed as 
$\epsilon_{\mu\nu\rho\sigma}[D_{\nu},[D_{\rho},D_{\sigma}]]=0$. 
By direct calculations, one gets
$[D_{\rho},D_{\sigma}]=-igG_{\rho\sigma}+[\partial_{\rho},\partial_{\sigma}]$, where the second commutator term of the partial derivative operators can not be discarded when the gauge field has a line singularity of the Dirac type~\cite{Dirac:1931}. 
Actually, it is the origin of VNABI as shown in the following.
The relation $[D_{\nu},G_{\rho\sigma}]=D_{\nu}G_{\rho\sigma}$ and the Jacobi identities  lead us to
\begin{eqnarray}
D_{\nu}G^{*}_{\mu\nu}=-\frac{i}{2g}\epsilon_{\mu\nu\rho\sigma}[D_{\nu},[\partial_{\rho},\partial_{\sigma}]]=
\frac{1}{2}\epsilon_{\mu\nu\rho\sigma}[\partial_{\rho},\partial_{\sigma}]A_{\nu}=\partial_{\nu}f^{*}_{\mu\nu}. \label{eq-JK}
\end{eqnarray}
 Eq.(\ref{eq-JK}) shows that VNABI $J_{\mu}=D_{\nu}G^{*}_{\mu\nu}$ is equivalent to that of the Abelian  Bianchi identities $k_{\mu}=\sum_{a=1}^3k_{\mu}^a\sigma^a/2=\partial_{\nu}f^{*}_{\mu\nu}$.
Eq.(\ref{eq-JK}) is gauge covariant and therefore a non-zero  $J_{\mu}$ is a gauge-invariant property. As shown in \cite{Suzuki:2014wya,Suzuki:2017lco},  $J_\mu=k_\mu$ transforms as an adjoint operator and satisfies both $D_\mu J_\mu=0$ and $\partial_\mu J_\mu=0$~\cite{Arafune:1975}.  There exist magnetic local $U(1)_m^3$ ( in $SU(2)$) and local $U(1)_m^8$ ( in $SU(3))$ symmetries kinematically in addition to the global color symmetry. As proved in \cite{Suzuki:2014wya,Suzuki:2017lco}, the Dirac quantization condition for each magnetic monopole is satified.

     VNABI exists  only when gauge fields in QCD contain a line singularity of the Dirac type~\cite{Dirac:1931}.      
This means that the Abelian monopoles from VNABI are completely different from 
those obtained in the Abelian projection scheme contrary to the expectation of \cite{Bonati:2010tz}, since the latter comes from the singularity of the gauge fixing matrix 
$V(x)$. 

Let us summarize advantageous points of VNABI as Abelian monopoles for all colors in the following:
\begin{enumerate}
\item \textit{Color confinement is proved in the framework of ADME.} \\
Although there are $U(1)_m^3$ symmetries in $SU(2)$ QCD, only one $U(1)_e$ is the maximal Abelian torus group. When the Abelian monopole $k_\mu^{a=3}$ make condensation for example, electric charged states with respect to $U(1)_e^{a=3}$ are confined. However, the state $(\bar{u}_3u_3-\bar{d}_3d_3)$
which belongs  to a $SU(2)$ triplet is not confined, since it is neutral with respect to $U(1)_e^{a=3}$. 
But it is possible to choose another $U(1)_e^{a=1}$ or $U(1)_e^{a=2}$ instead of $U(1)_e^{a=3}$.   The state $\bar{u}_3u_3-\bar{d}_3d_3$  is charged with respect to $U(1)_e^{a=1}$ and  is confined since the Abelian monopole $k_\mu^{a=1}(x)$  also make condensation. It is important to note that, when monopole condensation occurs in all three-color directions, physical states become $SU(2)$ color singlets alone. 
\item \textit{Perfect Abelian and monopole dominances for all color components without any additional gauge fixing.}\\
The color electric flux between a static quark and antiquark is squeezed which leads us to a linear potential with the string tension $\sigma$. If the flux squeezing is owing to a solenoidal Abelian monopole current as suggested by ADME, the string tension is explained by those of Abelian and monopole static potentials, i.e.,  $\sigma=\sigma_a=\sigma_m$. This is called as Abelian and monopole dominances.
It is possible to extract a link field $\theta_\mu^a$ having a color $a$ out of $U_{\mu}(s)$ both in $SU(2)$~\cite{Suzuki:2008} and $SU(3)$~\cite{Ishiguro:2022} and a lattice monopole~\cite{DeGrand:1980eq}. Perfect Abelian and monopole dominances 
 are seen numerically  even without any additional gauge-fixing~\cite{Suzuki:2008, Suzuki:2009,Ishiguro:2022,Suzuki:2023}. 
\item \textit{Existence of the continuum limit of Abelian monopoles (VNABI) for all color conponents.}\\
The renormalization-group (RG) study~\cite{Wilson:1974B} utilizing a block-spin transformation is a powerful method to study the continuum limit in the framework of lattice field theories~\cite{Aizenman:1982,Frohlich:1982}. 
The block-spin RG study is also very powerful to study the continuum limit of the monopole dynamics~\cite{Ivanenko:1991wt} described by integer Abelian lattice monopole variables~\cite{Shiba:1994db}. It is possible to prove the existence of the continuum limit of the monopole density and infrared effective monopole action~\cite{Swendsen:1984} in $SU(2)$~\cite{Shiba:1994db,Suzuki:2017lco,Suzuki:2017zdh} and in $SU(3)$~\cite{Suzuki:2023,Suzuki:2024}. 
\end{enumerate}

The assumption of  VNABI is beyond the usual framework of QCD and hence it is important to check the effect of VNABI in other areas of QCD besides color confinement.       
     The aim of this note is to study the effect of VNABI on the QCD topology such as topological charge and chiral $U(1)$ symmetry breaking
.
Section 2 deals theoretically with two  effects of VNABI on QCD topology. A new strange term  $L(x)=2\Tr(J_{\mu}(x)A_{\mu}(x))$ composed of the product of the gauge field and VNABI appears both in the relation between the winding number and the topological charge and in the chiral $U(1)$ anomaly relation.  However, $\Lambda=(g^2/16\pi^2)\int d^4xL(x)$  is shown to vanish theoretically.  Section 3 discusses numerically $\Lambda$ with the help of the gradient flow. The data are consistent with $\Lambda=0$ in the continuum limit. In Section 4, another biggest effect of VNABI on QCD topology is shown. It is the interesting fact that 
self-dual instantons can not be a classical solution of QCD 
at space-time points where VNABI occurs. One has to find an alternative mechanism explaining integer topological charge, etc. In addition, 
 a new interesting relation $Q_a$=3$Q_t$ between the topological charge and its Abelian conterpart is derived. This may become a clue to clarify the origin of the topological charge instead of instantons. The relation is numerically studied on lattice. Final section is devoted to summary and some remarks. 

\section{VNABI and topology in QCD}\label{Sec2}

\subsection{The topological charge}
The topological charge $Q_t(A)$ and the topological charge density are defined \cite{Mueller-Preussker:2011} by
\begin{eqnarray}
Q_t(A)=\frac{g^2}{16\pi^2}\int d^4x\rho_t(x),\ \ \rho_t(x)\equiv\Tr(G_{\mu\nu}G^*_{\mu\nu}) \label{TC}
\end{eqnarray}
and gauge-variant Chern-Simons density is written as
\begin{eqnarray*}
K_\mu(x)
=\frac{g^2}{8\pi^2}\epsilon_{\mu\nu\rho\sigma}\Tr
\left[A_\nu(x)(\partial_\rho A_\sigma(x)
-\frac{2ig}{3}A_\rho(x)A_\sigma(x))\right].  
\end{eqnarray*}
Evaluate the divergence $\partial_\mu K_\mu(x)$:
\begin{eqnarray}
\partial_\mu K_\mu(x)&=&\frac{g^2}{16\pi^2}[\rho_t(x)-L(x)],\ \ \ 
L(x)\equiv 2\Tr(A_\mu(x)J_\mu(x)).\label{JA1}
\end{eqnarray}
$L(x)$ is zero if gauge fields are regular as usually assumed. But when VNABI exists, it can be non-vanishing. 
 
\subsection{A vacuum-to-vacuum amplitude}
Let us compute a vacuum-to-vacuum amplitude:
\begin{eqnarray*}
Z&=&<vac|\exp(-\frac{1}{\hbar}\hat{H}(\tau-\tau_0)|vac>.
\end{eqnarray*}
The vacuum $A^{vac}_{\mu}(x)$ is characterized by 
an integer $\omega_{\infty}$
as follows:
\begin{eqnarray}
\omega_{\infty}=\oint d^3\sigma n_{\mu}K_{\mu} 
=\frac{ig^3}{24\pi^2}\oint d^3\sigma n_{\mu}\epsilon_{\mu\nu\rho\sigma}\Tr\left[A_{\nu}A_{\rho}A_{\sigma}\right], \label{omega1}
\end{eqnarray}
where at the boundary $G_{\mu\nu}=0$
 and then $\epsilon_{\mu\nu\rho\sigma}\partial_{\rho}A_{\sigma}=ig\epsilon_{\mu\nu\rho\sigma}A_\rho A_\sigma$. $A_{\mu}$ is described by a pure gauge $\Omega\partial_\mu\Omega^{\dag}/ig$, where $\Omega$ is a regular gauge transformation matrix. Thus
\begin{eqnarray}
\omega_{\infty} = \frac{1}{24\pi^2}\oint d^3\sigma
n_{\mu}\epsilon_{\mu\nu\rho\sigma}\Tr\left[(\Omega^{\dag}\partial_{\nu}\Omega)(\Omega^{\dag}\partial_{\rho}\Omega)(\Omega^{\dag}\partial_{\sigma}\Omega)\right],\label{omega2}
\end{eqnarray}
which is the
Pontryagin index 
taking an integer. Hence we get
\begin{eqnarray}
\omega_{\infty}=Q_t(A)-\Lambda,\ \ \  \Lambda \equiv \frac{g^2}{16\pi^2}\int d^4x L(x). \label{Winding} 
\end{eqnarray}

\if01   
  Consider a large gauge transformation $\hat{T}(U)$ such that $\hat{T}|n>=|n+1>$.
Hamiltonian $\hat{H}$ is invariant under $\hat{T}(U)$: $[\hat{T},\hat{H}]=0$. Physical gauge invariant $\theta$-vacuum is defined as 
$\hat{H}|\theta>=E_0|\theta>$ where 
$|\theta>\equiv\sum_{n=-\infty}^{n=\infty}e^{in\theta}|n>$.
The vacuum transition amplitude is calculated as
\begin{eqnarray*}
Z=<\theta'|\exp(-\hat{H}\tau)|\theta>
=\sum_{n',n}e^{i(n\theta-n'\theta')}\int DA_{\mu}(x)|_{n,n'}e^{-S[A]}.
\end{eqnarray*}
$\int DA_{\mu}(x)|_{n,n'}\exp(-S[A])$ depends on $\nu=n-n'$. 
From eq.(\ref{Winding}), we see $n'-n = Q_t(A) - \Lambda$ and then
\begin{eqnarray*}
 Z=\sum_{n}e^{in(\theta-\theta')}\sum_{\nu}\int DA_{\mu}(x)|_{\nu}
e^{-S[A]-i(Q_t(A)-\Lambda)\theta'}.
\end{eqnarray*}

\fi

\subsection{Chiral $U(1)$ anomaly and VNABI}
Let us next discuss the chiral $U(1)$ anomaly.
Here we follow the works~\cite{Fujikawa:1979, Alvarez-Gaume:1984}.
When we define $u_i(x)$ as the normalized zero-mode eigenvectors of the Dirac operator $\slashed{D}_{\mu}=(\partial_{\mu}+igA_{\mu})\gamma_{\mu}$
of which $n_{+}$ eigenvectors have positive chirality $\gamma_5u_i(x)=u_i(x)$ and $n_{-}$ have negative chirality $\gamma_5u_i(x)=-u_i(x)$. Then without VNABI, we get the Atiyah-Singer index theorem 
$n_{-}-n_{+}=Q_t(A)$~\cite{Fujikawa:1979}. 
When VNABI exists, what happens about the Atiyah-Singer index theorem? 

Consider the gauge invariant Lagrangian
\begin{eqnarray*}
L=\bar{\psi}(\slashed{D}+m)\psi+\frac{1}{2}TrG_{\mu\nu}G_{\mu\nu}
\end{eqnarray*}
and perform a flavor singlet $\gamma_5$ transformation
$\psi(x)\to \exp[i\alpha(x)\gamma_5]\psi(x)$.
Then we get for infinitesimal $\alpha(x)$
\begin{eqnarray*}
L \to L-i\partial_{\mu}\alpha(x)\bar{\psi}\gamma_{\mu}\gamma_5\psi(x)
-2im\alpha(x)\bar{\psi}\gamma_5\psi.
\end{eqnarray*}
We also take into account the change of the measure of the functional integral, i.e., the Jacobian factor. Expand $\psi(x)$ and $\bar{\psi}(x)$ as 
$\psi(x)=\sum_n a_n\phi_n(x),\ \ \ 
\bar{\psi}(x)=\sum\phi_n^{\dag}(x)\bar{b}_n$,
 in terms of a complete set of eigenfunctions of the Hermitian operator $\slashed{D}$: 
$\slashed{D}\phi_n(x)=\lambda_n\phi_n(x)$.
For infinitesimal $\alpha(x)$, the Jacobian factor becomes 
\begin{eqnarray*}
\exp\left(-2i\int d^4x\alpha(x)\sum_n \phi_n^{\dag}(x)\gamma_5\phi_n(x)\right),
\end{eqnarray*}
which is divergent and can be regularized with a Gaussian cut-off:
\begin{eqnarray*}
\sum_n \phi_n^{\dag}(x)\gamma_5\phi_n(x)
&\to&\lim_{M\to\infty}\sum_n
\phi_n^{\dag}(x)\gamma_5\exp[-(\lambda_n/M)^2]\phi_n(x)\\
&=&\lim_{M\to\infty, y\to x}\Tr \gamma_5\exp[-(\slashed{D}/M)^2]
\delta(x-y).
\end{eqnarray*}
Now 
\begin{eqnarray*}
(\slashed{D})^2&=&D_{\mu}D_{\mu}+\frac{1}{4}\left([\partial_{\mu},\partial_{\nu}]+gG_{\mu\nu}\right)[\gamma_{\mu},\gamma_{\nu}].
\end{eqnarray*}
When the gauge fields are regular,  the term $[\partial_{\mu},\partial_{\nu}]$ is dropped. But when VNABI exists, it cannot be discarded. Actually 
\begin{eqnarray*}
&&\Tr \gamma_5 e^{-(\slashed{D}/M)^2}\delta(x-y)\\
&=&e^{-\frac{D_{\mu}^2}{M^2}}\Tr \gamma_5 (\frac{1}{2!})(-\frac{1}{4M^2})^2[\gamma_{\mu},\gamma_{\nu}][\gamma_{\alpha},\gamma_{\beta}]([\partial_{\mu},\partial_{\nu}]+gG_{\mu\nu})([\partial_{\alpha},\partial_{\beta}]+gG_{\alpha\beta})).
\end{eqnarray*}
Then finally we get
\begin{eqnarray}
\sum_k\phi_k^{\dag}(x)\gamma_5\phi_k(x) 
=-\frac{g^2}{16\pi^2}Tr(G^{*}_{\mu\nu}G_{\mu\nu
})+\frac{g^2}{8\pi^2}\Tr(J_{\mu}A_{\mu}). \label{zeromode}
\end{eqnarray}
The lefthand side of eq.(\ref{zeromode}) vanishes for the eigenfunctions with eigenvalue $\lambda_k\neq 0$ since  $\gamma_5\phi_k(x)$ has the eigenvalue $-\lambda_k$
 and is orthogonal to $\phi_k(x)$. 
 Then LHS becomes 
\begin{eqnarray*}
\int d^4x\left(\sum_{i=1}^{n_{+}}\phi^*_{i+}(x)\phi_{i+}(x)- \sum_{i=1}^{n_{-}}\phi^*_{i-}(x)\phi_{i-}(x)\right)
=n_{+}-n_{-},
\end{eqnarray*}
where $\phi_{i\pm}(x)$ are zero-mode eigenfunctions with the chirality $\pm$.
Hence we get 
\begin{eqnarray}
n_{+}-n_{-} = -Q_t+\Lambda. \label{zeromode2}
\end{eqnarray}  
 
   Now when we compare (\ref{zeromode2}) with (\ref{Winding}), we see
\begin{eqnarray}
n_{-}-n_{+}=\omega_{\infty}.
\end{eqnarray}
The Atiyah-Singer index theorem is unchanged also under the existence of VNABI. 
Eq.(\ref{Winding}) and eq.(\ref{zeromode}) show that the modifications due to VNABI appear always as a combination of $Q_t-\Lambda$. All previous discussions about the topological charge $Q_t$ are replaced by $Q_t-\Lambda$. 

\subsection{Vanishing of the additional term $\Lambda$}\label{subsection2-5}
It is important to note that all terms in (\ref{Winding}) and (\ref{zeromode}) except the additional term $\Lambda$ are gauge-invariant and integer. However, $\Lambda$ does not look gauge invariant nor integer. Hence $\Lambda$ must be zero when VNABI could exist in QCD.  Here let us discuss if $\Lambda=0$ is proved theoretically.
\begin{enumerate}

\item The additional term $\Lambda=(g^2/16\pi^2)\int d^4x 2\Tr(J_{\mu}(x)A_\mu(x))$  is composed of the Abelian monopoles as VNABI and the gauge field $A_\mu(x)$. The gauge field has  a regular electric part and a singular magnetic part producing the magnetic monopole. Let us denote 
the former as $a^i_{\mu}(x)$ and the latter as $b^i_{\mu}(x)$ where the superscript $i$ stands for the color ($i=1\sim 3$) in $SU(2)$. It is natural to assume $A^i_\mu(x)=a^i_{\mu}(x)+b^i_{\mu}(x)$ as seen from the Abelian gauge field after an Abelian projection~\cite{tHooft:1981ht} below.
\item 
Let us first discuss the singular gauge field.
Wu-Yang~\cite{Wu-Yang:1975} considered Dirac's monopole in $U(1)$ electrodynamics in terms of a non-integrable phase factor. The gauge field describing the 
static 
monopole is defined on $R^3\backslash \{0\}$ or $R^4\backslash \{\textrm{world line}\}$ which are homotopy equivalent to $S^2$. We parametrize 
$S^2\ni (\hat{\Phi}_1, \hat{\Phi}_2, \hat{\Phi}_3)$ in terms
 of polar coordinates as $(\Phi_1=r\sin\theta\cos\phi,\Phi_2=r\sin\theta\sin\phi, \Phi_3=r\cos\theta)$ where $\theta\in (0,\pi)$ and $\phi\in [0,2\pi)$.
It is impossible to define a unique gauge field $A_{\mu}(x)$ on all $S^2$. Divide $S^2$ into two hemispheres $S^2_N\ (0\le \theta\le \pi/2$ and $S^2_S\ (\pi/2\le\theta\le\pi)$ and define gauge fields on each hemisphere as 
\begin{eqnarray}
A^N_{\mu}(x)&=&\frac{g_m}{4\pi}(1-\cos\theta(x))\partial_{\mu}\phi(x) \nn\\
&=&\frac{g_m}{4\pi}\frac{1}{1+\hat{\Phi}_3(x)}\epsilon_{3bc}\hat{\Phi}_b(x)\partial_{\mu}\hat{\Phi}_c(x)\label{AN}\\
A^S_{\mu}(x)&=&\frac{-g_m}{4\pi}(1+\cos\theta(x))\partial_{\mu}\phi(x)\label{AS},
\end{eqnarray}
where $\hat{\Phi}_a(x)\equiv\Phi_a(x)/r(x)$.
Only angle variables appear in (\ref{AN}) and (\ref{AS}).
The gauge field $A^N$ ($A^S$) has a line singularity at $\cos\theta(x)= -1\ \ (+1)$. Both give the Dirac monopole existing at the center. The above gauge fields (\ref{AN}) and (\ref{AS}) at the boundary of the two hemispheres are related by a a single-valued gauge transformation if the Dirac quantization condition $(2eg_m)/(4\pi)$=integer~\cite{Dirac:1931} is satisfied. The flexibility in the choice of the overlapping regions and in the choice of the gauge fields were also discussed in \cite{Wu-Yang:1975}.

\item 
Wu-Yang~\cite{Wu-Yang:1975} generalized the above discussions to one of three $U(1)$ subgroups in
 non-Abelian $SU(2)$. This is similar to the case of VNABI considered here. Since the minimum charge $g/2$ is one-half of the charge $g$ of the gauge field, the Dirac quantization condition becomes $(gg_m)/(4\pi)$=integer. Contrary to the $U(1)$ case,  it is enough to consider one candidate of the gauge field on all $S^2$.  Gauge fields having different directions of the Dirac string give rise to the same monopole field. We here consider a gauge field having a line-like singularity upto a regular gauge transformation
\begin{eqnarray}
b^i_{\mu}(x)&=&\frac{g_m}{4\pi}(1-\cos\theta^i(x))\partial_{\mu}\phi^i(x)\label{BV1}\\
&=&\frac{g_m}{4\pi(1+\hat{\Phi}_3^i(x))}\epsilon_{3bc}\hat{\Phi}_b^i(x)\partial_{\mu}\hat{\Phi}_c^i(x). \label{BV2}
\end{eqnarray}
There is a line singularity at $1+\hat{\Phi}_3^i(x)=0$ in this case.
The flexibility of the choice of the gauge field in this case was also discussed in \cite{Wu-Yang:1975}.

\item
The field strengths and the monopole currents from the singular gauge fields~(\ref{BV1}) are written by
\begin{eqnarray}
f^i_{\mu\nu}(x)
&=&\frac{g_m}{4\pi}\sin\theta^i(x)(\partial_{\mu}\theta^i(x)\partial_{\nu}\phi^i(x)-\partial_{\nu}\theta^i(x)\partial_{\mu}\phi^i(x)
)
\nn\\
&=&\frac{g_m}{4\pi}\epsilon_{abc}\hat{\Phi}^i_a(x)\partial_{\mu}\hat{\Phi}^i_b(x)\partial_{\nu}\hat{\Phi}^i_c(x)\nn\\
k^i_{\mu}(x)&=&\partial_{\nu}f^{i*}_{\mu\nu}(x)=\frac{g_m}{8\pi}\epsilon_{\mu\nu\rho\sigma}\epsilon_{abc}\partial_{\nu}\hat{\Phi}^i_a(x)\partial_{\rho}\hat{\Phi}^i_b(x)\partial_{\sigma}\hat{\Phi}^i_c(x).\label{KV}
\end{eqnarray}
Eq.(\ref{BV2}) and eq.(\ref{KV}) lead us kinematically to 
\begin{eqnarray}
\int d^4x \sum_{i=1}^3\sum_{\mu}b^i_{\mu}(x)k^i_{\mu}(x)=0.\label{AK3}
\end{eqnarray}
The singular gauge fields $b^i_{\mu}(x)$ do not contribute to $\Lambda$. 
In the case of a static monopole, (\ref{AK3}) is trivial, since 
$b^i_4(x)=0$ and $k^i_4$ alone is non-zero.

\item
It is to be noted that a similar singular gauge field appears in the case of the 'tHooft Abelian projection~\cite{tHooft:1981ht} when a gauge transformation $V(x)$ of $SU(2)$ into $U(1)$ is done.   Any type of Abelian projection is specified by a Higgs-like scalar field
$\hat{\psi}_3(x)=V(x)^{\dag}\sigma_3V(x)=\sum_{a=1}^3\hat{\psi}_3^a(x)\sigma^a$ transforming as an adjoint operator. After performing such a partial gauge fixing,  the  induced Abelian gauge field is written by 
\begin{eqnarray}
A'_{\mu}(x)&=&V(x)A_{\mu}(x)V^{\dag}(x)-\frac{i}{g}\partial_\mu {g}V(x)V^{\dag}(x)\nn\\
&\to& A_{\mu}^{'3}(x)=\Tr A'_{\mu}(x)\sigma_3=a^3_\mu(x)+b^3_\mu(x)\label{A31}\\
b^{3}_\mu(x)&=&\frac{1}{g(1+\hat{\psi}_3^3(x))}\epsilon_{3ab}\hat{\psi}_3^a(x)\partial_{\mu}\hat{\psi}_3^b(x)=(1/g)(1-\cos\theta_3(x))\partial_\mu\phi_3(x),\label{A3}
\end{eqnarray}
where $b^{3}_\mu(x)$ is a singular part of the induced gauge field, $a^3_{\mu}(x)=\Tr\hat{\psi}_3(x)A_{\mu}(x)$ is regular. Here  
$\hat{\psi}_3^1(x)=\sin\theta_3(x)\cos\phi_3(x), \hat{\psi}_3^2(x)=\sin\theta_3(x)\sin\phi_3(x), \hat{\psi}_3^3(x)=\cos\theta_3(x)$. Also 
\begin{eqnarray}
k_{\mu}^3(x)&=&\partial_{\nu}f^{*}_{\mu\nu}(x)=\frac{1}{2}\epsilon_{\mu\nu\rho\sigma}\epsilon_{abc}\partial_{\nu}\hat{\psi}_3^a(x)\partial_{\rho}\hat{\psi}_3^b(x)\partial_{\sigma}\hat{\psi}_3^c(x),
\end{eqnarray}
so that $\int d^4x \sum_{\mu}b^{3}_{\mu}(x)k^3_{\mu}(x)=0$ similarly. 

\item 
Now how about the contribution from the regular gauge fields $a^i_{\mu}(x)$ to $\Lambda$ ? It is to be noted that the regular fields satisfy the Abelian Bianchi identities
\begin{eqnarray}
\epsilon_{\mu\nu\rho\sigma}\partial_{\nu}\partial_{\rho}a^i_{\sigma}(x)=0.\label{aB}
\end{eqnarray}
The contribution from the regular fields to $\Lambda$ is given by
\begin{eqnarray*}
&&\int d^4x \sum_{i} a^i_{\mu}(x)\epsilon_{\mu\nu\rho\sigma}\partial_{\nu}\partial_{\rho}b^i_{\sigma}\\
&=&\int d^4x\sum_i\epsilon_{\mu\nu\rho\sigma}
\partial_{\nu}(a^i_{\mu}(x)\partial_{\rho}b^i_{\sigma}(x))
-\partial_{\rho}(\partial_{\nu}a^i_{\mu}(x)b^i_{\sigma}(x),
\end{eqnarray*} 
where eq.(\ref{aB}) is used. 
There are only the contributions from the boundary. At the boundary,
 the non-Abelian field strength should 
vanish since otherwise the action diverges. Then a gauge field is described by a pure gauge, i.e., $a_{\mu}(x)\to \Omega(x)\partial_{\mu}\Omega^+(x)/ig$ and $b_{\mu}(x)\to 0$. 
When
 we write (\ref{BV1}) for the static monopole in the rectangular coordinates, we see $b_1(x)$ and $b_2(x)$ behave as $1/r$ 
when $r\to \infty$  and $b_3(x)=b_4(x)=0$.  The contribution from the regular gauge fields to $\Lambda$ is also vanishing. Hence the additional term $\Lambda$ is expected to vanish theoretically. Numerical studies of $\Lambda$ will be done in the next section.
\end{enumerate}

\section{Lattice numerical studies of QCD topology and $\Lambda$}\label{Sec3}
Although naively lattice gauge fields are topologically trivial, the discussion of the topology on a lattice is meaningful sufficiently close to the continuum limit~\cite{Luscher:1982}. 
On a lattice, gauge fields are described by a compact field 
$U(s,\mu)$
 on a link $(s,\mu)$ where $s$ denotes a site and $\mu$ is a direction. As a lattice gauge action,  we adopt the following:
\begin{eqnarray}
    S =   \beta (c_0\sum_{pl} S_{pl}
       +  c_1\sum_{rt} S_{rt}), \label{Action}
\end{eqnarray}
where $S_{pl}$ and $S_{rt}$ denote a $1 \times 1$ plaquette and $1 \times 2$ rectangular loop terms in the action. Here we adopt  the tadpole-improved action where $c_0=1$ and $c_1=-1/(20u_0)^2$ and $u_0$ is the $SU(2)$ tadpole improvement factor~\cite{Alford:1995hw}.  The lattice size adopted is $24^4$. We consider $\beta=3.0\sim 3.7$ and the lattice spacings are between from 
$a(3.0)=0.17279(19)$fm to  $a(3.7)=0.05252(23)$fm.  For comparison, we also consider the usual Wilson action with $c_0=1\ \ c_1=0$
 at $a(2.3)=0.16634(167)$fm $\sim$ $a(2.65)=0.0529(19)$fm.
\begin{figure}[htb]
\begin{minipage}{0.45\textwidth}
\caption{$\beta$ dependence of the minimum, the maximum and the averaged absolute value of $\Lambda$ among 50 configurations in the tadpole-improved action under MCG and MAU1 gauge fixings.\label{MMA-Lambda-TI}}
  \centering
  \includegraphics[width=\linewidth]{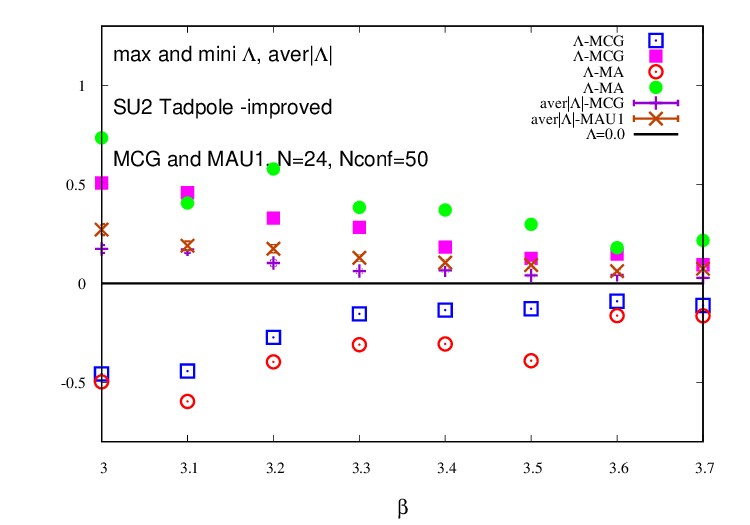}
\end{minipage}
\hfill
\hspace*{.5cm}
\begin{minipage}{0.45\textwidth}
\caption{$\beta$ dependence of the minimum, the maximum and the averaged absolute value of $\Lambda$ among 50 configurations in the Wilson action under MCG  gauge fixing.\label{MMA-Lambda-W}}
  \centering
  \includegraphics[width=\linewidth]{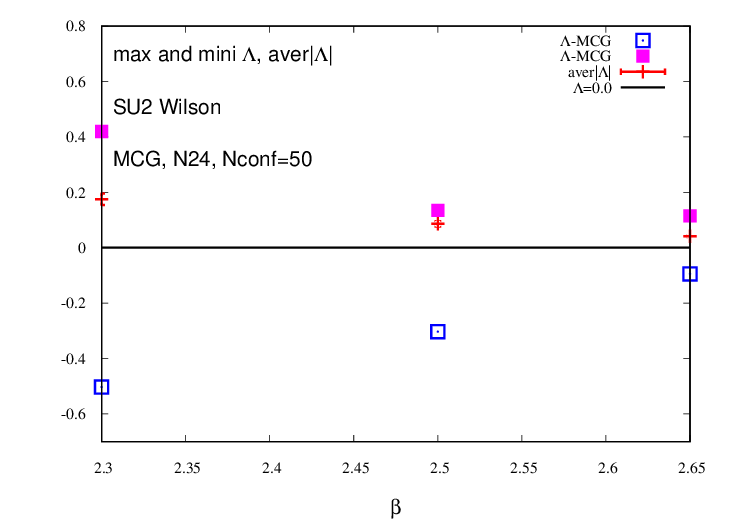}
\end{minipage}

\vspace*{1cm}
\end{figure}

\subsection{Evaluation of the additional term $\Lambda$}
Let us consider what happens about the additional term $\Lambda\equiv g^2\int d^4x2\Tr(A_{\mu}(x)J_{\mu}(x))/(16\pi^2)$ on the lattice. 
Since the lattice QCD at finite lattice spacing on a finite volume is finite, VNABI defined naively in terms of $U_{\mu}(s)$ alone always vanish. However since
VNABI $J_{\mu}$ in $\Lambda$ is  written  in terms of Abelian monopole currents $k^a_{\mu}(x)$ as in eq.(\ref{eq-JK}), the latter of which can be defined on lattice following DeGrand-Toussaint~\cite{DeGrand:1980eq}. 

\begin{enumerate}
  \item     First of all, we have to extract a link field $\theta_\mu^a$ having a color $a$ out of $U_{\mu}(s)$. The following method is found to be applicable for Abelian confinement pictures  both in $SU(2)$~\cite{Suzuki:2008} and $SU(3)$~\cite{Ishiguro:2022} in accord with the previous definition used in the old Abelian projection.
We maximize the following quantity locally
\begin{eqnarray}
R^a= \mathrm{Re} \tr\left\{\exp(i\theta_\mu^a(s)\lambda^a)U_\mu^{\dag}(s)\right\}, \label{RA}
\end{eqnarray}
where $\lambda^a$ is the Gell-Mann matrix and the sum over $a$ is not taken. In $SU(2)$, eq.(\ref{RA}) leads us to
\begin{eqnarray}
\theta_\mu^a(s) = \tan^{-1}\left\{\frac{U_\mu^a(s)}{U_\mu^0(s)}\right\},\label{theta2}
\end{eqnarray}
where $U_\mu(s)=U_\mu^0(s)+ i\sum_a\sigma^aU_\mu^a(s)$. In the calculation of $\Lambda$, we  consider a compact form $sin\theta_\mu^a(s)$ as a field corresponding to $A_\mu^a(x)$ in $\Lambda(x)$. 
\item
Next at present a reliable lattice monopole definition is the one proposed by DeGrand-Toussaint~\cite{DeGrand:1980eq} which counts the number of the Dirac string coming out of monopoles. 
 A plaquette variable 
$\theta_{\mu\nu}^a(s)\equiv \partial_\mu\theta_\nu^a(s)-\partial_\nu\theta_\mu^a(s)$ where $\theta_\mu^a(s)$ is an Abelian link field of color $a$. It is decomposed into 
$\theta_{\mu\nu}^a(s)= \bar{\theta}_{\mu\nu}^a(s) +2\pi n_{\mu\nu}^a(s)$, 
 where $\bar{\theta}_{\mu\nu}^a(s)\in [-\pi,\pi]$. 
 Then the integer $n_{\mu\nu}^a(s)\in [-2,2]$ 
could be regarded as a number of the Dirac string penetrating the plaquette, although the number can take $[-\infty,\infty]$ in the continuum theory. Then DGT~\cite{DeGrand:1980eq} defined a monopole current on a lattice as 
\begin{eqnarray}
k_{\mu}^a(s) = -\frac{1}{4\pi}\epsilon_{\mu\nu\rho\sigma}
\partial_{\nu}\bar{\theta}_{\rho\sigma}^a(s+\hat{\mu}). 
\label{dgcur} 
\end{eqnarray}
Since the Dirac quantization condition~\cite{Dirac:1931} is essential for existence of Abelian monopoles in QCD, we adopt here in  $SU(2)$ the above lattice monopole definition~\cite{DeGrand:1980eq} for each global color  and  study the continuum limit.  However there appear many lattice artifact monopoles which do not have the continuum limit. It is absolutely necessary to reduce lattice artifact monopoles as much as possible by adopting various smoothing methods: 

\begin{itemize}
\item
  Monopole currents satisfy the conservation condition. Hence lattice monopoles exist as closed loops on a lattice. It is already known that they are composed of one or a few long infrared (IF) (percolating) loops and many short loops (for example, see \cite{Ivanenko:1990, Kitahara:1995}).
 Only the former is known to play an important role in color confinement in the continuum limit. Hence it is important to adopt only the infrared monopoles and to study their behaviors.
\item Introduction of an additional partial gauge fixing is known to be powerful in reducing lattice artifact monopoles.  Here we adopt mainly the Maximal Center gauge (MCG)~\cite{DelDebbio:1996mh,DelDebbio:1998uu}. The so-called direct Maximal Center gauge~\cite{Faber:2001zs} which requires maximization of the quantity
\begin{eqnarray}
R=\sum_{s,\mu}(\tr U_{\mu}(s))^2 \label{eq:MCG}
\end{eqnarray}
with respect to local gauge transformations.  The condition (\ref{eq:MCG}) fixes the gauge up to $Z(2)$ gauge transformation and can be considered as the Landau gauge for the adjoint representation. Hence MCG does not belong to an Abelian projection where the maximal torus group $U(1)$ remains unbroken~\cite{tHooft:1981ht}.  In our simulations, we choose simulated
annealing algorithm as the gauge-fixing method which is known to be  very powerful for finding the global maximum. For details, see the reference~\cite{Bornyakov:2000ig}.
MCG violates $SU(2)$ but keeps global $SU(2)$ symmetry.  We also considered the Maximal Abelian gauge (MAG)~\cite{Kronfeld:1987ri,Kronfeld:1987vd} which is known also as a gauge fixing reducing lattice artifact monopoles. MAG is the gauge which maximizes
$R=\sum_{s,\hat\mu}{\rm Tr}\Big(\sigma_3 U_{\mu}(s)\sigma_3 U^{\dagger}_{\mu}(s)\Big) \label{R}$
with respect to local gauge transformations.  But MAG breaks $SU(2)$ into the maximal torus group $U(1)$ violating also global $SU(2)$ symmetry. In MAG alone, there remain a lot of lattice artifact monopoles with respect to  off-diagonal components. Hence we also do additional Landau gauge fixing about the remaining $U(1)$ symmetry after MAG (this is called as MAU1). Under these gauge fixings, the density of  infrared monopoles are found also to be small. 
\end{itemize}
\end{enumerate}

The numerical results of  $\Lambda(s)$ are  fluctuating largely around zero as shown in the tadpole-improved action under the MCG and the MAU1 gauges (figure \ref{MMA-Lambda-TI}) and in the Wilson action under the MCG gauge  (figure \ref{MMA-Lambda-W}) respectively for 50 configurations. The figures show that $\Lambda(s)$   
seems
non-zero but decreases as $\beta$ go to larger thus approaching to the continuum limit $\beta\to\infty$. What is the continuum limit is not known yet from these analyses alone. 
\begin{figure}[htb]
\begin{minipage}{0.45\textwidth}
  \caption{The additional term $\Lambda$ in the tadpole-improved action with MCG gauge  under the gradient flow}\label{Lambda_flow_TIMCG}
  \centering  
  \includegraphics[width=\linewidth]{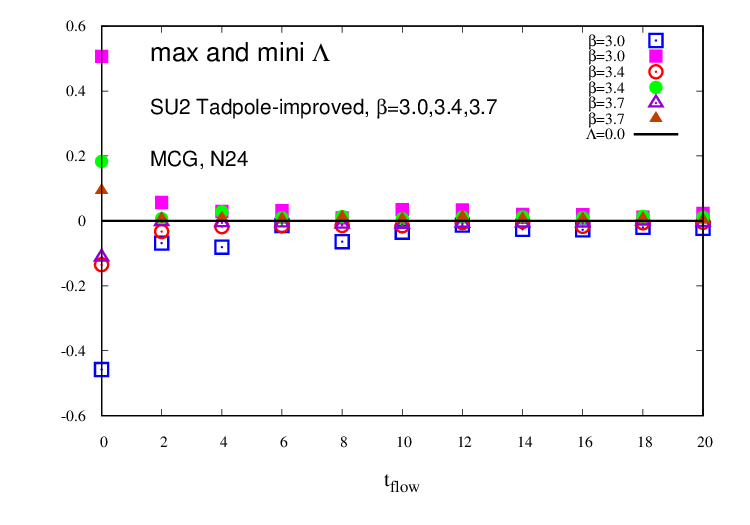}
\end{minipage}
\hfill
\hspace*{.5cm}
\begin{minipage}{0.45\textwidth}
  \caption{The additional term $\Lambda$ in the Wilson action with MCG gauge  under the gradient flow}\label{Lambda_flow_WMCG}
\centering   
  \includegraphics[width=\linewidth]{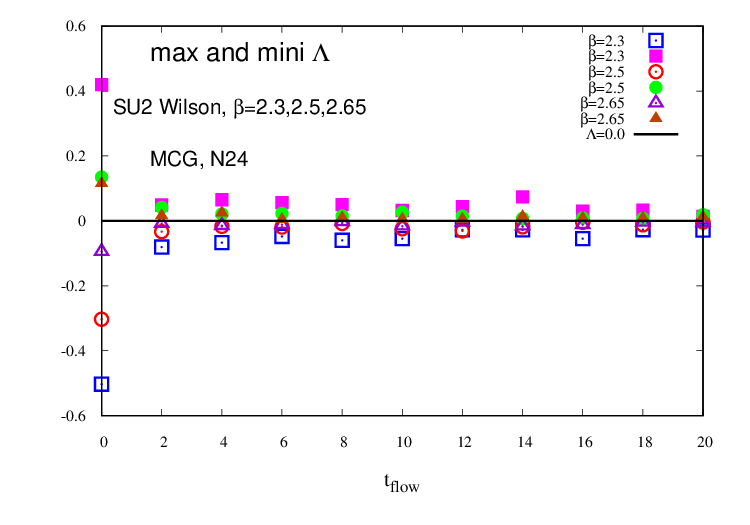}
\end{minipage}
\end{figure}

\subsection{Introduction of the gradient flow} 
Let us next study what happens after suppressing ultraviolet fluactuations by means of the gradient flow.  The gradient flow ( and equivalently a controlled cooling method ) is known to be powerful for reducing ultraviolet 
fluctuations \cite{Luscher:2011}. As the flow action, we adopt DBW2 action~\cite{DBW2:2000} where $c_0=1-8c_1,\ \ c1=-1.4088$ in (\ref{Action}) following the recent work~\cite{Tanizaki:2024} on  the stability of the topological charge  under the gradient flow. Here all gradient flows are calculated by using the third-order Runge-Kutta method with the time step $\epsilon=0.01$~\cite{Luscher:2010}.

Adopting three $\beta$ points as typical examples having similar lattice distances both in the tadpole-improved and Wilson actions, we plot the maximum and minimum $\Lambda$ at the flow time upto $t_{flow}=20$ as shown in 
figure \ref{Lambda_flow_TIMCG} and figure \ref{Lambda_flow_WMCG}. 
It is found that $\Lambda$  becomes almost zero  above around 
$t_{flow}=2$
 both in the improved action (figure \ref{Lambda_flow_TIMCG}) and the Wilson action  (figure \ref{Lambda_flow_WMCG}) under MCG gauge fixing, although in MAU1, more unstable behaviors are seen in comparison with those in MCG.

To be noted that 
$\Lambda$ tends to vanish rapidly after the gradient flow in every case.
The rapid decrease of $|\Lambda(\beta)|$ as $\beta$ becomes larger and  
the rapid drop under the gradient flow  strongly suggest that $\Lambda$  vanishes in the continuum limit.

\section{Monopoles and instantons}\label{Sec4}
\subsection{Correspondence between instantons and monopoles}
It is known that a classical solution of Euclidian QCD called instanton~\cite{Belavin:1975} is important in the quantum transition between degenerate vacua with different winding numbers. Instantons give us also an integer topological charge $Q_t$. The instanton is a solution satisfying the self-duality condition $G_{\mu\nu}=\pm G^{*}_{\mu\nu}$. 

Instantons and monopoles are two typical topological objects in quantum field theories.
There are various works suggesting the intimate relations between  monopoles and 
instantons. In QCD, for example, Kondo~\cite{Kondo:2026} showed recently that $D=3$ magnetic monopoles are constructed from symmetric instantons in $d=4$ Euclidian Yang-Mills theory. Numerically there are also many works suggesting the existence of deep relations between Abelian monopoles and 
instantons~\cite{Giacomo-Hasegawa:2015,Hasegawa:2022,Hart-Teper:1995,Sasaki:1996}.  
\subsection{Instantons in the presence of VNABI}
However it is stressed that, when VNABI exists and plays an important role in QCD color confinement,  such an instanton solution can not be allowed
 at the space-time points where VNABI exists. Actually
 the self-duality condition of instantons leads us automatically to the non-Abelian Bianchi identity 
$D_{\mu}G^{*}_{\mu\nu}=0$
from the equation of motion of QCD, that is, 
$D_{\mu}G_{\mu\nu}=0$. 

Even when we consider an instanton solution only at space points without VNABI, it can not explain the integer topological charge, since such a solution does not give us 
$|Q_t(A)|=1$ for one instanton case. Note that $Q_t(A)$ is given by an integral over all space points, i.e., $Q_t\equiv (g^2/16\pi^2)\int d^4x \Tr(G_{\mu\nu}G_{\mu\nu}^*)$. 
How much difference occurs depends on how densely VNABI distributes in all space points. 
Numerically  VNABI exists in the continuum limit forming one or a few long percolating closed loops in four-dimensional Euclidian space and condensation of Abelian monopoles as VNABI occurs as shown in \cite{Suzuki:2017lco,Suzuki:2017zdh,Suzuki:2023,Suzuki:2024}. It is known that configurations of closed monopole loops are expressed in terms of Abelian dual Higgs model~\cite{Suzuki:1988,Maedan:1989,Koma:2003C} containing a monopole scalar field. When the monopole condensation occurs, the monopole scalar field has a vacuum expectation value. This suggests that instantons can not exist at any space point when monopole condensation occurs.   

Hence when VNABI exists, one has to find another new mechanism which could solve phenomena such as  integer topological charges and chiral symmetry breaking, etc, which are usually expected to be solved by instantons. This may be the biggest effect of VNABI on QCD topology.
A few comments are in order.
\begin{enumerate}
\item This work is the first showing the incompatibility of instantons and monopoles in a theoretically rigorous way. All works cited above do not assume such a singularity in gauge fields producing VNABI.
  \item It is needless to say that VNABI can not say any classical solution without the self-duality or anti self-duality condition.
\end{enumerate}
\begin{figure}[htb]
\begin{minipage}{0.45\textwidth}
\caption{$Q_t$, $Q_a$ and $\Lambda$ versus gradient flow time in a configuration at $\beta=3.5$ in MCG.\label{Qt_Qa_Lambda_MCG}}

\vspace*{.5cm}
  \centering
  \includegraphics[width=\linewidth]{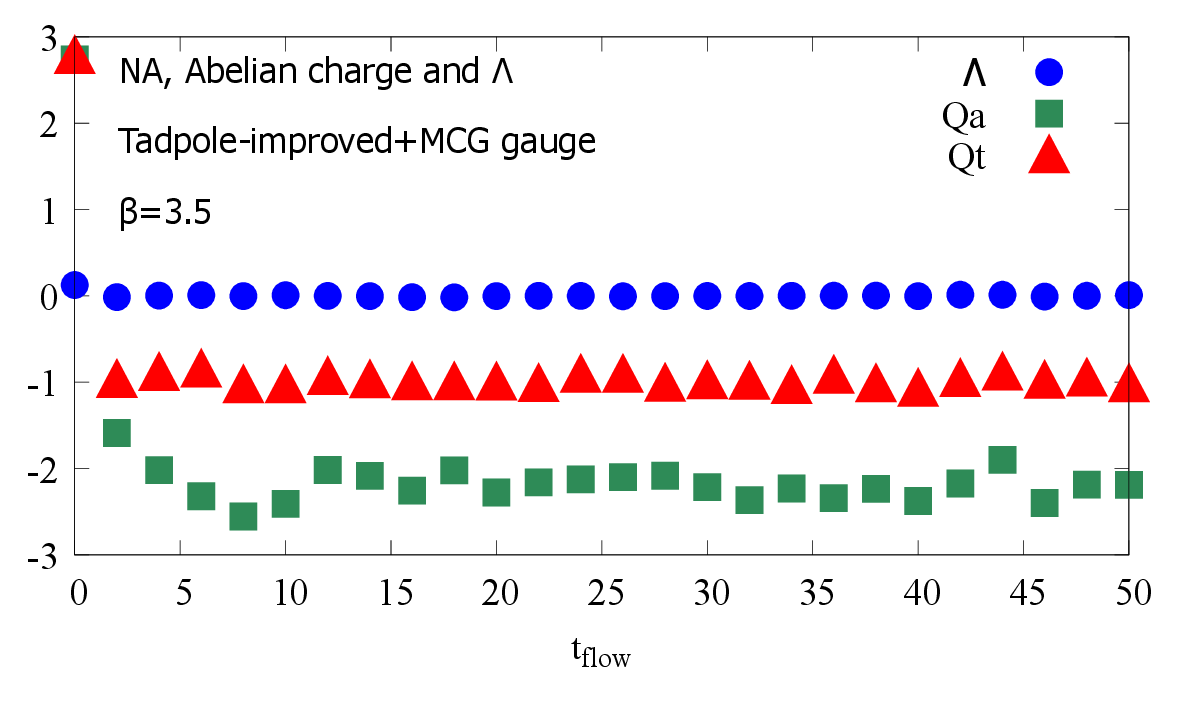}
\end{minipage}
\hfill
\begin{minipage}{0.45\textwidth}
\caption{$Q_t$, $Q_a$ and $\Lambda$ versus gradient flow time in a configuration at $\beta=3.5$ in MAU1.\label{Qt_Qa_Lambda_MA}}

  \vspace*{.5cm}
  \centering
  \includegraphics[width=\linewidth]{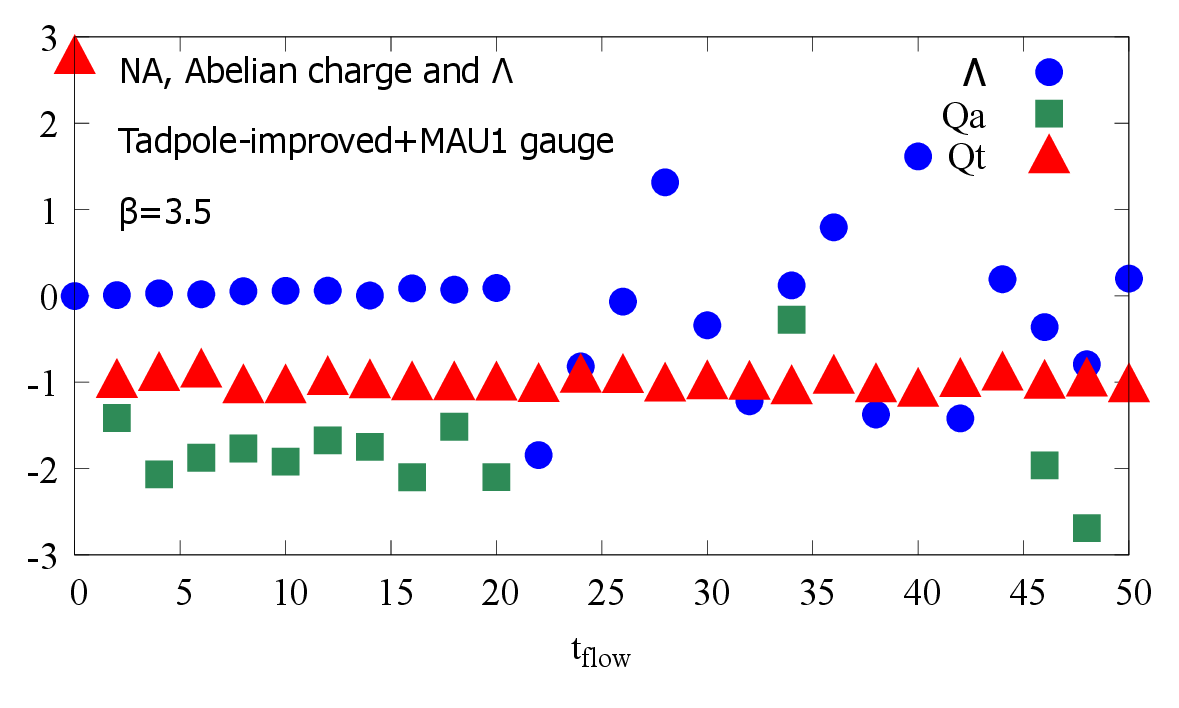}
\end{minipage}
\end{figure}

\subsection{An Abelian mechanism behind the QCD topology}
In order to find an alternative mechansim in place of instantons, there may exist any Abelian mechanism as suggested below.
\begin{enumerate}
\item
Since $k_{\mu}(x)=\partial_{\nu}f^{*}_{\mu\nu}(x)$, it is possible to prove 
\begin{eqnarray}
\Lambda&=&\frac{g^2}{8\pi^2}\int d^4x \Tr(\partial_{\nu}f^*_{\mu\nu}(x)A_{\mu}(x))\nn\\
&=&\frac{g^2}{8\pi^2}\int d^4x \partial_{\nu}\Tr(f^*_{\mu\nu}(x)A_{\mu}(x)) + \frac{g^2}{16\pi^2}\int d^4x \Tr(f_{\mu\nu}f^*_{\mu\nu})\nn\\
&=&Q_a(A)+\frac{g^2}{8\pi^2}\int d^4x n_{\nu}\epsilon_{\mu\nu\rho\sigma}\Tr(\partial_{\rho}A_{\sigma}(x)A_{\mu})\nn\\
&=&Q_a(A)-\frac{ig^3}{8\pi^2}\int d^4x n_{\mu}\epsilon_{\mu\nu\rho\sigma}\Tr(A_{\mu}A_{\rho}A_{\sigma})\nn\\
3\omega_{\infty}&=&Q_a(A)-\Lambda, \label{AWinding}\\
Q_a(A)&\equiv& \frac{g^2}{16\pi^2}\int d^4x \Tr(f_{\mu\nu}f^*_{\mu\nu}), \nn
\end{eqnarray}
where we use the same boundary condition $G_{\mu\nu}(x)\to 0$ again and eq.(\ref{omega1}) and eq.(\ref{omega2}). 
To be noted that $Q_t(A)$ is equal to the integral over an inner product of non-Abelian electric and magnetic fields, whereas
$Q_a(A)$ is that of Abelian electric and magnetic fields. It is interesting to see that when $\Lambda=0$, we get $Q_a(A)=3Q_t(A)$ and 
Abelian $Q_a(A)$ is also gauge invariant. 
This sugggests that the Abelian magnetic and electric fields in the presence of $J_{\mu}$ condensation could have a role in explaining the topological charge 
$Q_t$ in place of instantons.
However, the present author can not yet underatand at present why $Q_a(A)$ takes an integer value. 
\item 
The relation $Q_a(A)=3Q_t(A)$ is similar to the Abelian dominance of the string tension. The non-Abelian string tension $\sigma$ is regarded as an average over the string tensions $\sigma_a^i\ \ (i=1\sim 3)$ of the three colored Abelian fluxes in $SU(2)$. When the global color symmetry is kept, $\sigma_a^1=\sigma_a^2=\sigma_a^3$. The Abelian dominance says $\sigma=\sigma_a^i$ and so $\sum_i \sigma_a^i=3\sigma$. 
To be noted that, when we define an Abelian topological charge for each color component as $Q^i_a(A)\equiv (g^2/16\pi^2)\int d^4x f^i_{\mu\nu}f^{i*}_{\mu\nu}$ and use the fact that $\Lambda=0$ holds good for each color component separately as shown in \ref{subsection2-5}, it is possible to prove directly that $Q^1_a(A)=Q^2_a(A)=Q^3_a(A)=Q_t(A)$ in $SU(2)$. However, in $SU(N)$ for $N\ge 3$, the Pontryagin index is written by a sum of integers~\cite{Christ:1980} and hence such  simple relations between $Q^i_a(A)$ as in $SU(2)$ do not hold.

\item 
With respect to  the topological charge $Q_t$, two ways of definitions are used so far~\cite{Alexandrou:2020}, i.e., the fermionic definition using (\ref{zeromode}) and the bosonic definition using (\ref{Winding}). The former is always integer when the lattice Dirac operator satisfies the Ginsparg-Wilson relation~\cite{GW:1982} as  in the overlap Dirac operator~\cite{Neuberger:1998}. However to get integer values from the bosonic definition is not so easy. Various improvements are needed to get a stabilized value as shown for example in ~\cite{Alexandrou:2020,Tanizaki:2024}. Here we choose the DBW2 action~\cite{DBW2:2000} following  \cite{Tanizaki:2024}, since it seems best to get a stable $Q_t$ rapidly. As the definition of the lattice $Q_t$, here we adopt those composed of $1\times 1$ Wilson loops for simplicity. Adopting only 10 configurations at each $\beta$, only one configuration at $\beta=3.5$ gives us a stabilized $Q_t$. Hence we pay attention to the configuration and study the flow-time history up to $t_{flow} =50$ as done in ~\cite{Tanizaki:2024}. 
\item
We measure the Abelian candidate of the topological charge $Q_a$ obtained from (\ref{AWinding}).
On the lattice, we adopt an Abelian $1\times 1$ plaquette variable as the compact form 
$f^L_{\mu\nu}(s)=\sin(\theta_{\mu\nu}(s))$. See figure \ref{Qt_Qa_Lambda_MCG}.
Although non-Abelian $Q_t$ becomes stabilized when $t_{flow}>2$, the Abelian $Q_a$ are fluctuating between -2 and -3 in MCG gauge. It should become -3 theoretically when the definition would be correct. In MAU1 gauge case, $Q_a$ is smaller around -2. Moreover, very unstable behaviors start above $t_{flow} >20$ with respect to both $Q_a$ and $\Lambda$ as seen in figure \ref{Qt_Qa_Lambda_MA}. 
More elaborate calculations on much larger lattices for larger $\beta$ must be needed along with an additional  introduction of smoothing techniques like the block-spin transformation~\cite{Shiba:1994db,Suzuki:2017lco,Suzuki:2017zdh} and improved definitions of $Q_a$ must be done.
\end{enumerate}
\section{Summary and remarks}\label{Sec5}
In this note, it is studied how VNABI affects  QCD topological features besides color confinement mechansim. As a whole, it is found that VNABI could exist in QCD without causing serious problems. The Atiyah-Singer index theorem and the chiral $U(1)$ anomaly which are related to the topological charge are found to be unchanged. The biggest effect seems to be that (anti) self-dual instantons can not exist at the space-time points where VNABI exists. Hence instantons can not explain integer topological charge nor the chiral $U(1)$ anomaly. We have to find another mechanism behind such phenomena. The  Abelian counter term $Q_a(A)$ which is a product of Abelian electric and magnetic field strengths is found to satisfy $Q_a(A)=3Q_t(A)$. This may be a clue.

Finally some remarks are in order.    
\begin{enumerate}
\item
    Lattice numerical studies~\cite{Suzuki:2017lco,Suzuki:2017zdh,Suzuki:2023}  suggest the existence of the continuum limit of Abelian monopoles due to VNABI. However it is known that  gauge fields giving Abelian monopoles have a Dirac string with a line singularity and field strengths have a point-like singularity~\cite{Dirac:1931}. There must be a mechanism in QCD therefore admitting existence of Abelian monopoles in the continuum. The problem was discussed already in the works~\cite{Chernodub:2001A,Bornyakov:2002,Chernodub:2003A}. With respect to the Dirac string, it cost no action on the lattice formulated in terms of a compact link field. They showed that it is possible to extend the continuum naive QCD action in such a way to allow for the Dirac strings without any cost in the action. With respect to the point-like singularity, the entropy factor of the monopoles is important in addition to the energy. Compatibility with the asymptotic freedom of QCD needs a fine-tuning between the energy and the entropy. The numerical data on the lattice seem to show the fine-tuning. 
Our discussions here assume that a similar mechanism is working and topology of QCD can be discussed under the existence of VNABI.
\item
 What happens about  the usual chiral $SU(2)\times SU(2)$ symmetry breaking without instantons? There are works suggesting a strong correlation between Abelian magnetic monopoles and the chiral 
 condensate~\cite{Ohta:2021,Suganuma:2021}. 
\item
When the gauge field has a line-like singularity in $QED$, there can exist a monopole called Dirac monopole~\cite{Dirac:1931}. However, such a monopole is not found experimentally~\cite{ATLAS:2024}.
We propose VNABI as Abelian magnetic monopoles assuming the existence of a Dirac-type singularity in non-Abelian gauge fields. To search for the existence of such Abelian monopoles in experiments are very important. The spontaneously broken dual $U(1)_m^8$ predicts existence of color singlet scalar bosons and axial vector dual gauge bosons which are  eightfold degenerate with respect to the global color components.
It is necessary to analyse ADME in dynamical $SU(3)$ QCD to analyse properties including masses of these new bosons. 
\item
Monopoles are expected to play a key role in confinement-deconfinement transition of the universe. It is very interesting to study the mechanism and the role of the monopoles at $T>T_c$ is also very interesting in the framework of dynamical full QCD (See a preliminary work~\cite{Bornyakov:2005}).
\end{enumerate}

\section*{Acknowledgements} 
 The numerical simulations of this work were done  using High Performance Computing resources at Research Center for Nuclear Physics  (RCNP) of Osaka University.  The author would like to thank  RCNP for their support of computer facilities. 

\end{document}